\documentclass{article}
\usepackage{graphicx}
\begin{document}
\author{\"{O}zg\"{u}r Delice
\footnote {Department of Physics, Bogazici University, 34342
Bebek, Istanbul, Turkey; e-mail: odelice@boun.edu.tr}}

\title{
Kasner generalization of Levi-Civita spacetime}
 \maketitle

\begin{abstract}{
We investigate some cylindrically symmetric nonstationary and
nonstatic solutions of Einstein field equations. We first study
some physical properties of a solution which can be considered as
Kasner generalization of static Levi-Civita vacuum solution. Then
we generalize this metric to include a solution where a space-time
filled with null dust or a stiff fluid. }
\end{abstract}

\section{Introduction}
One of the most important differences of the spherically and the
cylindrically symmetric  vacuum solutions of General Relativity is
that, according to the Birkhoff theorem, there is a timelike
Killing vector in the spherically symmetric vacuum solution. Thus,
it can be said that the spherically symmetric vacuum is
necessarily static.
 However, the situation drastically changes when we consider the
cylindrically symmetric systems since there is no analogue of
Birkhoff's theorem in cylindrical symmetry. During the
gravitational collapse of a cylindrically symmetric system,
gravitational waves can be emitted  and the exterior region of a
collapsing cylindrical body is not static \cite{thorne}. This fact
has important consequences in the studies of gravitational waves,
cosmological models, quantum gravity and numerical relativity.

If $\partial_z $ and $\partial_\phi$ are the axial and the angular
Killing vectors describing  cylindrical symmetry, then the most
general cylindrically symmetric nonstationary metric  can be
written in the canonical form as \cite{kramer}:
\begin{equation}\label{weyl}
ds^2=e^{2(K-U)}(-dt^2+dr^2)+e^{2U}dz^2+e^{-2U}W^2d\phi^2,
\end{equation}
where $K,U$ and $W$ are the functions of $r$ and $t$ in general.
Here $r$ is the radial, $t$ is the time, $z$ is the axial and
$\phi$ is the angular coordinate with the ranges $0\le r <
\infty$, $-\infty<z,t<\infty$, $0\le \phi \le 2 \pi$.

The static solution of the metric (\ref{weyl}) represents the
exterior gravitational field of a general static cylindrical line
source and was found by
  Levi-Civita in 1919 \cite{levicivita}:
 \begin{equation}
ds^{2}=-\rho ^{4s }dt^{2}+\rho ^{4s (2s -1)}(d\rho
^{2}+ dz^{2})
+\alpha^2\rho ^{2(1-2s )}d\phi^{2}, \label{Lc}
 \end{equation}
where here $s$ and $\alpha$ are constant parameters. These
parameters in general cannot be removed by a coordinate
transformation if $\phi$ is an angular coordinate. Research on
this solution was mainly focused on the understanding of these
parameters and finding physically acceptable sources generating
this metric since in order to understand the meaning and the
behaviour of the metric parameters $s$ and $\alpha$, one may need
to match it with an interior solution. Static cylinders
\cite{cylinders} and cylindrical shells \cite{Shells} have been
constructed as a source of this metric. Shells composed of various
matter sources satisfying some energy conditions  for certain
ranges of $s$ have also been studied \cite{photonic}. The
parameter $s$ is related to the energy density of the source and
$\alpha$ is an angular deficit parameter.

The nonstatic vacuum solutions of (\ref{weyl}) have also been
studied extensively. They have important consequences on
cosmology, gravitational waves and also on quantum gravity. For a
discussion of these solutions we refer to the book of Stephani et.
al. \cite{kramer}.

Furthermore, the time dependent cylindrically symmetric nonvacuum
solutions of Einstein equations were studied for different
cylindrical systems. An expanding cylindrical radiation filled
universe \cite{davidson}, a radiation universe with heat and null
radiation flow \cite{patel}, nonstatic cosmic strings with a time
dependent vacuum exterior \cite{Schabes,Shaver,Som}, nonstatic
global strings \cite{Sen} are some examples of such solutions.
Some of these solutions have an interesting property that their
exterior vacuum solutions correspond to some particular values of
the parameters the Levi-Civita metric having also a Kasner type
time dependence. This fact motivates us to study the Kasner
generalization of the Levi-Civita solution with the full range of
its parameters. Thus in this paper, we will study the properties
of cylindrically symmetric time dependent vacuum solutions in
Kasner form. This solution can be considered as a Kasner
generalization of the Levi-Civita (LC) solution since for every
constant time slice it reduces to the LC solution. These kind of
generalized Kasner solutions having more than one variable are
well known and studied by different authors \cite{Mcintosh}. This
solution is also equivalent to the Einstein-Rosen soliton  wave
solutions \cite{Carmeli} by a coordinate transformation. Although
this solution is well known, we will establish a direct relation
between the parameters of the LC solution with the parameters of
its Kasner generalization. We will also perform a detailed
comparison of the LC solution and its nonstatic Kasner
generalization by studying their singularity behaviour, geodesic
structure and radial acceleration of test particles in these
spacetimes.

We also extend our discussion into some nonvacuum  generalizations
of this solution. Since the gravitational collapse of a physically
reasonable source is one of the main topics in general relativity,
the cylindrical collapse is studied extensively in the literature
\cite{thorne,Shapiro}. The radiating Levi-Civita space-time
\cite{krishna}, a space-time filled with a radially oriented null
radiation in an otherwise empty static background is generally
employed in these concerns and others \cite{Pereira}, since this
metric can represent the exterior region of a collapsing
cylindrical body. However, since there is no analogue of Birkoff
theorem in cylindrical symmetry, it might be reasonable to discuss
a nonstatic generalization of this radiating Levi-Civita solution.
We also present a  nonstatic stiff fluid  as an another example of
this form.

 The paper is organized as follows.
In the next section we present the Kasner generalization of the LC
solution. In the section (III) we discuss some physical properties
of this solution. The Section (IV) discusses the radiating
generalization of this solution and its some physical properties.
In the Section (V) we present a solution representing a universe
filled with a nonstatic isotropic stiff fluid.  Lastly, we give a
brief conclusion.

\section{Levi-Civita-Kasner Solution}
Let us consider the following ansatz for the functions of the
metric (\ref{weyl}):
\begin{eqnarray}\label{klcansatz}
&& W=\alpha (c_1r+c_2)(c_3t+c_4),\\
&& U=k\ln (c_1r+c_2)+q\ln(c_3t+c_4),\\ &&
K=k^2\ln(c_1r+c_2)+q^2\ln(c_3t+c_4),
\end{eqnarray}
where $k,\ q,\ \alpha$ and $ c_i$'s  are constants. Here, when
$c_3=0$, $c_4\neq 0$, $c_1 \neq 0$ we get the Levi-Civita solution
of the form:
\begin{equation}\label{lc1}
ds^2=r^{2(k^2-k)}(-dt^2+dr^2)+r^{2k}dz^2+\alpha^2r^{2(1-k)}d\phi^2
\end{equation}
where we have rescaled the coordinates $r,t$ and $z$.  One can
recover
the conventional form of the LC solution (\ref{Lc}) by applying 
the following coordinate transformations:
\begin{equation}\label{N}
R=\frac{r^\kappa}{\kappa},\ R=\frac{\rho^S}{S},\ \kappa=k^2-k+1,\
S=4s^2-2s+1,\ s=\frac{k}{2(k-1)}.
\end{equation}

 When we choose $c_1=0$, $c_2 \neq 0$,
$c_3 \neq 0$ the solution reduces to well known vacuum Kasner
solution:
\begin{equation}
ds^2=t^{2(q^2-q)}(-dt^2+dr^2)+t^{2q}dz^2+t^{2(1-q)}d\phi^2,
\end{equation}
where $q$ is a real constant. For this case the coordinates can be
thought of as the Cartesian coordinates. The coordinate
transformation {$t'\,=\,(Q\,t){Q^{-1}}$} puts the Kasner solution
in its familiar form  as \cite{Kasner}:
\begin{equation}
ds^2=-dt^2+t^{2a}dr^2+t^{2b}dz^2+t^{2c}d\phi^2,
\end{equation}
where we have again rescaled the metric, removed prime for clarity
and $a=(q^2-q)Q^{-1}$, $b=q \, Q^{-1}$, $c=(1-q)Q^{-1}$ with
$Q=q^2-q+1$. Kasner solution corresponds to an anisotropic
homogenous cosmology. Here the constants $a,b,c$ satisfy the
Kasner constraints:
\begin{equation}a+b+c=1=a^2+b^2+c^2.
\label{kasconst}
\end{equation}

Also, for $c_1=c_3=0$ and others nonvanishing we get flat
Minkowski spacetime. Notice that $c_1$ and $c_2$ cannot vanish
simultaneously in (\ref{klcansatz}). Same is true also for $c_3$
and $c_4$.

If one calculates the Ricci tensor of the metric
(\ref{klcansatz}),
  the only nonvanishing term is:
\begin{equation}\label{r01}
R_{01}=-c_1 c_3(-1+ k^2 - 2kq +q^2)(c_1r+c_2)^{(-1
+2k-2k^2)}(c_3t+c_4)^{(-1+2q-2q^2)}.
\end{equation}
Here we see that when $c_1$ or $c_3$ vanish we have a vacuum
solution as it should be. Assuming they do not vanish,  equating
(\ref{r01}) to zero we get $q=k\pm 1$ which results (Hereafter we
choose $c_2=c_4=0$ and we absorb $c_1$ and $c_3$ in the
coordinates $r,t,z$ by redefining them):
\begin{eqnarray}\label{lck}
ds^2&=&r^{2(k^2-k)}\, t^{2((k+\epsilon)^2-k-\epsilon)}(-dt^2+dr^2)
\nonumber \\&& + r^{2k}\, t^{2(k+\epsilon)}dz^2 +\ P^2\,
r^{2(1-k)}\, t^{2(1-k-\epsilon)}d\phi^2,
\end{eqnarray}
with $\epsilon=\pm1$. Thus, we have  obtained the desired Kasner
generalization of the LC solution, where we can call it as
Levi-Civita-Kasner space-time (LCK). It is better to express them
with the Levi-Civita parameter since we want to compare them with
the static solution. The transformations:
\begin{equation}
 R=r^{\kappa}\,\kappa^{-1},\ \tau=Q^{-1}t^{Q},\ k=2s/(2s-1),\
 Q=(k+\epsilon)^2-(k+\epsilon)+1,
\end{equation}
leads to the following metric:
\begin{eqnarray}\label{lckGN}
ds^2=-R^{2D}d\tau^2+\tau^{2A}dR^2+R^{2E}\tau^{2B}dz^2+\alpha^2\,R^{2F}\tau^{2C}d\phi^2,
\end{eqnarray}
where we again rescaled the coordinates  $\tau,R,z$, absorbed all
constant into $\alpha$ and
\begin{eqnarray}
&& H=\epsilon(4s^2-1)+(1-2s)^2,\quad A=\frac{2s+H}{S+H},\\
&&B=\frac{(2s-1)(2s+\epsilon(2s-1))}{S+H},\quad
C=\frac{(1-2s)(1+\epsilon(2s-1))}{S+H},
\\&& D=\frac{2s}{S},\quad
E=\frac{2s(2s-1)}{S},\quad F=\frac{1-2s}{S}.\phantom{AAAA}
\end{eqnarray}
 For any value of $s$ we have in general two different solutions depending on
 $\epsilon=\pm1$. These solutions are in the form of the generalized
 Kasner spacetimes \cite{Mcintosh} and the metric functions $A,B,C$ and $E,F,G$
  satisfy the Kasner constraints  separately:
\begin{eqnarray}
A+B+C=A^2+B^2+C^2=1,\\
D+E+F=D^2+E^2+F^2=1.
\end{eqnarray}

The LCK solution (\ref{lck}) is also equivalent to Einstein-Rosen
soliton wave
solutions \cite{Carmeli,kramer} by  a transformation: 
\begin{eqnarray}
r&=&\sqrt{T-\sqrt{T^2-\varrho^2}},\quad
t=\sqrt{T+\sqrt{T^2-\varrho^2}}
\end{eqnarray}
which puts the metric functions into the form:
\begin{eqnarray}
W&=&rt=\varrho\\
U&=&k\ln{\varrho}+\frac{\epsilon}{2}\ln\left(T+\sqrt{T^2-\varrho^2}\right)\\
K&=&k^2\ln \varrho+\left(\epsilon
k+\frac{1}{2}\right)\ln\left(T+\sqrt{T^2-\varrho^2}\right)-\frac{1}{2}\ln(\left(
2\sqrt{T^2-\varrho^2} \right)
\end{eqnarray}
This transformation is valid only for $t^2>r^2.$ For $r^2>t^2$ we
need the following transformation:
\begin{eqnarray}
r&=&\sqrt{\varrho+\sqrt{\varrho^2-T^2}},\quad
t=\sqrt{\varrho-\sqrt{\varrho^2-T^2}},
\end{eqnarray}
which gives:
\begin{eqnarray}
W&=&rt=T,\\
U&=&k\ln T+\frac{\epsilon}{2}\ln\left(
\varrho-\sqrt{\varrho^2-T^2}\right),\\
K&=&k^2\ln T -\left(\epsilon k +\frac{1}{2} \right)\ln
\left(\varrho-\sqrt{\varrho^2-T^2} \right)-\frac{1}{2}\ln\left(
2\sqrt{\varrho^2-T^2}\right).
\end{eqnarray}
Here the  first metric is not valid at $\varrho>T$ and the other
is not valid at $\varrho<T$. Then we need to extend one to join
with the other. After achieving this, the resulting spacetime is
the solution we consider in this article. Hence, the metric we
discuss covers both regions.
\section{Some Physical properties of LCK solution}
\subsection{NP Spin and Weyl Coefficients} The static Levi-Civita
metric is Petrov type I in general except it is  flat for
$s=0,1/2$ and it is Petrov type D for $s=-1/2,1/4,1$ (See da Silva
et al in \cite{cylinders}). Let us compare  with LCK spacetime.

Here, using a NP tetrad, we will present the nonvanishing spin and
Weyl scalars of this spacetime since in this formalism some of the
curvature components have direct physical meaning \cite{NP}. The
canonical form of the metric (\ref{lck}) is more appropriate for
our purposes. We chose the NP tetrad as follows:
\begin{eqnarray}
&&\phantom{}ds^2=\mathbf{l}\otimes \mathbf{n}  -\mathbf{m }\otimes
\bar{\mathbf{m}}, \\&&
\sqrt{2}\,\mathbf{l}=\mathbf{e}^0+\mathbf{e}^1,
\sqrt{2}\,\mathbf{n}=\mathbf{e}^0-\mathbf{e}^1,\
\sqrt{2}\,\mathbf{m}=\mathbf{e}^2+i\mathbf{e}^3,\\
&& \mathbf{e}^0=r^{k^2-k}t^{(k+\epsilon)^2-k-\epsilon}dt,\
\mathbf{e}^1= r^{k^2-k}t^{(k+\epsilon)^2-k-\epsilon}dr,\\
&& \mathbf{e}^2=r^kt^{k+\epsilon}dz,\ \mathbf{e}^3=\alpha
r^{1-k}t^{1-k-\epsilon}d\phi.
\end{eqnarray}
For $\epsilon=1$ the nonvanishing components of spin coefficients
and Weyl scalars are:
\begin{eqnarray}
&&\sigma=-\frac{(1+2k)\,r+(1-2k)\,t}{2\sqrt{2}r^{k^2-k+1}t^{k^2+k+1}},\quad
\lambda=\frac{(1+2k)\,r-(1-2k)\,t}{2\sqrt{2}\,r^{k^2-k+1}\,t^{k^2+k+1}},\\
&& \rho=\frac{t-r}{2\sqrt{2}\,r^{k^2-k+1}\,t^{k^2+k+1}},\quad
\mu=\frac{t+r}{2\sqrt{2}\,r^{k^2-k+1}\,t^{k^2+k+1}},\\
&&\epsilon=\frac{k((1+k)\,r+(1-k)\,t)}{2\sqrt{2}\,r^{k^2-k+1}\,t^{k^2+k+1}},\quad
\gamma=\frac{k((1-k)\,t-(1+k)\,r)}{2\sqrt{2}\,r^{k^2-k+1}\,t^{k^2+k+1}},\\
&&\kappa=\nu=\tau=\pi=\alpha=\beta=0,\\
&&\Psi_0=\frac{k\left((1+k)(1+2k)\,r^2+4(1-k^2)\,r\,t+(1-k)(1-2k)\,t^2\right)}{2\,r^{2k^2-2k+2}\,t^{2k^2+2k+2}},\\
&&\Psi_2=\frac{k((1+k)\,r^2+(1-k)\,t^2)}{2\,r^{2k^2-2k+2}\,t^{2k^2+2k+2}},\\
&&\Psi_4=\frac{k\left((1+k)(1+2k)\,r^2-4(1-k^2)\,r\,t+(1-k)(1-2k)\,t^2\right)}{2\,r^{2k^2-2k+2}\,t^{2k^2+2k+2}}.
\end{eqnarray}
This shows us that the LCK spacetime with $\epsilon=1$ is again
Petrov type I in general except it is flat for $k=0(s=0)$ and
$k\rightarrow \infty\ (s=1/2)$. Also, since $\kappa=0$,
$\mathbf{l}$ is geodesics but it is not affinely parameterized
since $\epsilon\neq 0$ except $k=0(s=0)$. It also has expansion
($-\rho\neq 0$) and shear ($|\sigma|\neq 0$) but it is not
twisting.

For $\epsilon=-1$ we have
\begin{eqnarray}
&&\sigma=\frac{(3-2k)\,r+(2k-1)\,t}{2\sqrt{2}\,
r^{k^2-k+1}\,t^{k^2-3k+3}},\quad
\lambda=\frac{(2k-3)\,r+(2k-1)\,t}{2\sqrt{2}\,
r^{k^2-k+1}\,t^{k^2-3k+3}},\\
&& \rho=\frac{t-r}{2\sqrt{2}\, r^{k^2-k+1}\,t^{k^2-3k+3}},\quad
\mu=\frac{t+r}{2\sqrt{2}\, r^{k^2-k+1}\,t^{k^2-3k+3}},\\
&&\epsilon=\frac{(k-1)((k-2)r-k\,t)}{2\sqrt{2}\,
r^{k^2-k+1}\,t^{k^2-3k+3}},\quad
\gamma=\frac{(k-1)((2-k)r-k\,t)}{2\sqrt{2}\,
r^{k^2-k+1}\,t^{k^2-3k+3}},\\
&&\kappa=\nu=\tau=\pi=\alpha=\beta=0,\\
&&\Psi_0=\frac{(k-1)((6-7k+2k^2)\,r^2-4k(k-2)\,r\,t+k(2k-1)\,t^2)}{2\,
r^{2k^2-2k+2}\,t^{2k^2-6k+6}},\\
&&\Psi_2=\frac{(k-1)((k-2)\,r^2-kt)}{2\,
r^{2k^2-2k+2}\,t^{2k^2-6k+6}},\\
&&\Psi_4=\frac{(k-1)((6-7k+2k^2)\,r^2+4k(k-2)\,r\,t+k(2k-1)\,t^2)}{2\,
r^{2k^2-2k+2}\,t^{2k^2-6k+6}}.
\end{eqnarray}

Thus, LCK with $\epsilon=-1$ is also Petrov type I in general
except $k=1\ (s\rightarrow\infty)$  and $k\rightarrow \infty
(s=1/2)$ where it is flat. Again, the vector $\mathbf{l}$ is
geodesics but not affinely parameterized except $k=1$. Also, it
has nonvanishing expansion and shear but it is not twisting. Thus,
the LC and LCK solutions have common Petrov types in general, but
they differ for some particular values of the parameter $s.$

\subsection{Singularity Behaviour}
 The Kretschmann scalar $\mathcal{K}=R_{abcd}R^{ abcd}$
of the metric (\ref{lckGN}) are
\begin{eqnarray}
\mathcal{K}&=& 64\, {s^2}\,{{(1-2s)}^2} \Bigg(\frac{{{(1-4
s)}^2}{r^{-8 s/(1-2s+4s^2)}}}{{{(1-6 s +12 s^2)}^3}\, {t^4}} +
\frac{t^{8 s(1-4 s)/(1-6 s+12 s^2 )}}{{{(1-2 s+4
{s^2})}^3\,{r^4}}} \nonumber
\\
&&\phantom{}-\frac{2\,{{(1-4 s)}^2}\, {r^{-4 s/(1-2 s+4 {s^2})}}
\,{t^{4 s(1-4 s) /(1-6 s +12 s^2)}}}{{{(1-2 s+4 {s^2})}^2}
{{(1-6 s +12 s^2)}^2\,r^2\,t^2}}\Bigg),\ (\epsilon=1), \phantom{}\label{kresch1} \\
\mathcal{K}&=& 64\,(1-2s)^2 \Bigg( \frac{(s-1)^2\ {r^{-8 s/(1-2 s
+ 4{s^2})}}}{(3-6s+4s^2)^3 t^4} + \frac{s^2\, {t^{{4 {{(1-2
s)}^2}}/{(3-6 s+4 {s^2})}}}}{(1-2s+4s^2)\, r^4 \,t^4}
\nonumber \\
&& \phantom{} -\frac{8 s^2(s-1)^2 {r^{4 s/(1-2s+4s^2)}} t^{2 (1-2
s)^2/(3-6s+ 4s^2 )}}{(3-6s+4s^2)(1-2s+4s^2)r^2t^4} \Bigg) ,\
(\epsilon=-1).
\end{eqnarray}

We see that, this spacetime has singularities as in the static
case. It is well known that the static Levi-Civita spacetime is
singular at $r=0$, except for $s=0$, $s=\pm 1/2$ and $s\rightarrow
\infty$. For these values of $s$, the solution is regular and
flat. If one compares the static solution with the nonstatic
solutions, one realizes that there are similarities and
differences. For $\epsilon=1$ only when $s=0$ or $s=1/2$, the
solution is locally flat. For other cases, there are
singularities. For $s=1/4$ we have a singularity at $r=0$ whereas
for $s\rightarrow \infty$ we have singularity at $t=0$. For all
other values of $s$ we have singularities at both $r=0$ and $t=0$.
When $\epsilon=-1$ the situation is also different.  For this case
when $s=0$ the solution is not locally flat but contains a
singularity at $t=0$. There are locally flat solutions when
$s=1/2$ or $s\rightarrow \infty$. For $s=1$ we have a singularity
at $r=0$. For other values of $s$ we have both line and big-bang
singularities.

  As we have mentioned, for $s=0$ and $s=1/2$ static Levi-Civita solution (\ref{Lc}) is flat.
Corresponding solutions for $\epsilon=\pm1$ are:
\begin{eqnarray}
ds^2&=&-d\tau^2+dr^2+\tau^2dz^2+\alpha^2r^2d\phi^2,\quad (s=0,\ \epsilon=1),\label{lck10}\\
ds^2&=&-d\tau^2+\tau^{4/3}dR^2+\tau^{-2/3}dz^2+\alpha^2R^2\tau^{4/3}d\phi^2,
\quad (s=0,\ \epsilon=-1),\phantom{AA}\label{lck11} \\
ds^2&=&-R^2d\tau^2+\tau^2dR^2+dz^2+\alpha^2 d\phi^2,\quad (s=1/2,\
\epsilon=\pm 1). \label{lckrindler}
\end{eqnarray}

Here the first and third metrics are flat whereas the second one
is curved. The first and third metrics can be put into standard
Minkowski form with a suitable coordinate transformation. The
first solution is presented in \cite{Schabes} as a possible
nonstatic exterior solution corresponding to a nonstatic string.
Also, the second solution (\ref{lck11}) was introduced in
\cite{patel,Som,Sen} as an exterior vacuum solution to their
interior  nonstatic stringlike cylindrical source. It is not
singular at $r=0$ but has a big bang singularity at $t=0$ since
its Kretschmann scalar is $\mathcal{K}\sim t^{-4}$. Thus, for some
specicif values of parameters, the LCK solution reduces to some
previously known solutions.

\subsection{Radial acceleration of test particles}

The radial acceleration of a free test particle at rest in the
coordinate system of (\ref{lckGN}) is given by:
\begin{equation}
\frac{d^2R}{d\tau^2}=-\frac{D}{R\,T^{2A}}.
\end{equation}
The radial acceleration in the static Levi-Civita spacetime can be
found by taking $A=0$. For the Levi-Civita spacetime, when the
parameter $s$ is positive, the axis is attractive and when $s$ is
negative, the axis is repulsive. For a particle in a constant
radius, the magnitude of the acceleration is increasing with
increasing $s$ when $0<s<1/2$, and decreases with increasing $s$
when $s>1/2$. For $s=0$, no radial force is exerted on a particle
at rest. For nonzero $s$, when the radial distance increases,
radial acceleration decreases.

\begin{figure}[h]
\begin{center}  
\includegraphics[width=11cm]{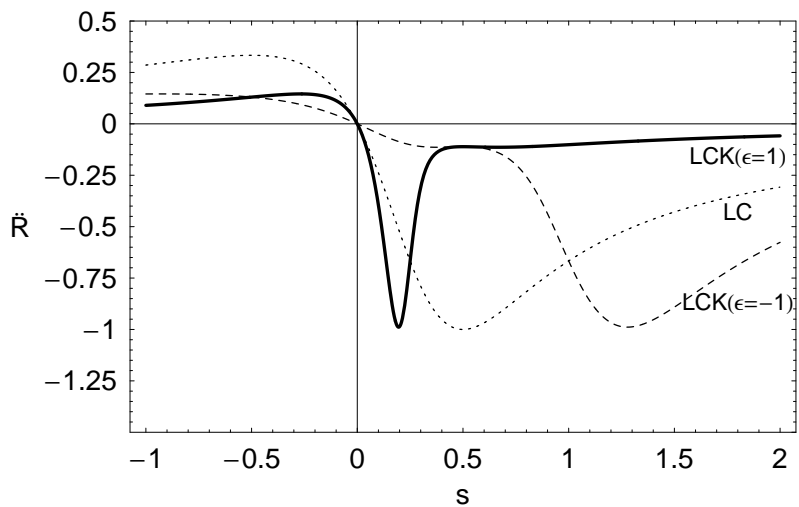}\\
\end{center}
\caption{The radial acceleration of a particle at $r=1$, $t=3$ for
the static Levi-Civita and Levi-Civita-Kasner spacetimes with
$\epsilon=\pm 1$. Dotted line represents static Levi-Civita
spacetime, the solid line represents the solution with
$\epsilon=1$ and dashed line represents $\epsilon=-1$ case. }
\label{racc}
\end{figure}

As in the Levi-Civita spacetime, for the Levi-Civita-Kasner
spacetimes (\ref{lckGN}) when $s$ positive the axis is attractive
and when $s$ negative the axis is repulsive. And also when $s=0$,
no radial acceleration is felt by a particle at rest. However,
since the solution is time dependent, the behaviour of
acceleration is changing with time. A typical behavior for $t=3$
can be seen in Figure \ref{racc}. Here, for $\epsilon=1$ the
magnitude of acceleration increases with increasing $s$ up to
$s\sim 0.17$, then it starts to decrease sharply up to $s\sim
0.3$, and then it decreases monotonically with increasing $s$. For
$\epsilon=-1$ situation is different. It increases monotonically
up to $s=1/2$ then increases more sharply up to $s=3/2$ then
starts to decrease with increasing $s$. When the time evolves, the
radial acceleration is getting stronger for certain ranges of $s$
and out of this range, particle feels very tiny force. For
$\epsilon=1$ this range is in between $0$ and $0.2$. For $\epsilon
=-1$ the situation is reverse. For small $s$ particle feels very
small force. The region where acceleration is very strong is near
$s\sim 1$. For other values of $s$ a test particle feels very tiny
radial force on it.

An important difference between LC and LCK spacetimes is that for
$s=1/2$ the radial acceleration of test particles becomes maximum
for LC metric. This fact has been discussed in previous studies of
LC metric since when the parameter $s$ increases the energy
density increases for $0<s<1/2$ but decreases for $s>1/2$. This
fact suggested that the parameter $s$ is somehow related with the
energy density of the source but not proportional to it. For the
LCK spacetime the maximum value of the radial acceleration is
different than $1/2$ (Fig. (\ref{racc})). Thus they have different
gravitational fields.

\subsection{Circular geodesics} Here we study  the equations of a
test particle following a circular geodesics in the spacetime
(\ref{lckGN}). The circular geodesics in the LC spacetime is
discussed in detail by da Silva et. al.\cite{cylinders}.

 Let us denote the angular velocity of a particle moving along a geodesics as
$\omega=d\phi/d\tau$ and its tangential velocity as
$W^{\mu}=(0,0,0, W^{\phi})$ with $W^{\phi}=\omega/\sqrt{-g_{tt}}$,
then we have (here $\tau$ is the time coordinate, not the proper
time and dot represents derivation with respect to an affine
parameter $\eta$):
\begin{eqnarray}
&&\left(\frac{ds}{d\tau}\right)^2=-R^{2D}+R^{2F} \tau^{2C}
\left(\frac{d\phi}{d\tau}\right)^2, \\
&&\omega^2=\left(\frac{\dot{\phi}}{\dot{\tau}}\right)^2=\frac{D}{\alpha^2F}
R^{2(D-F)}\tau^{-2C},\\
&&\ddot{\tau}=-C\alpha^2 R^{2(F-D)} \tau^{2C-1}\dot{\phi}^2,
\\ &&\dot{r}=0,\quad \dot{z}=0.
\end{eqnarray}

Then,
\begin{eqnarray}
&&W^{2}=\frac{D}{F}.
\end{eqnarray}
Replacing this into the first and the third equations, we get
\begin{eqnarray}
&&\left(\frac{ds}{d\tau}\right)^2=\left(W^2-1 \right)
R^{2D},\\
&& \dot{\tau}=\frac{d\tau}{d\eta}=\frac{\tau_0}{\tau^{C\sqrt{W}}}.
\end{eqnarray}
Thus, the circular geodesics are timelike for $W<1$ $(s<1/4)$,
spacelike for $W>1$ $(s>1/4)$ and null for $W=1$ (s=1/4). We have
the same conditions with the static Levi-Civita spacetime. Thus
the time dependence does not affect the circular geodesics. Also,
as in the static case, for a given $s$ the tangential velocity of
a particle is constant. The only difference between LC and LCK
spacetimes that is $\partial_\tau $ is not a Killing vector for
LCK spacetimes. This does not affect the dependence of the
character of the circular geodesics to the parameter $s$, although
they have different gravitational fields, since $\dot{\tau}$ is
not constant for this metric and also since the previous section
suggests.

\section{Radiating Levi-Civita-Kasner space-time}
It is well known \cite{krishna1,kramer} that for any
Einstein-Rosen wave solution with $(K=K_0,\ U=U_0,\ W=W_0)$
solving vacuum Einstein equations for this metric, there is a
corresponding radiative solution $(K=F(r-t)+K_0,\ U=U_0,\ W=W_0)$
satisfying:
\begin{equation}\label{emt}
T_{\mu\nu}=\eta\, \mathbf{k}_\mu \mathbf{k}_\nu
\end{equation}
where $\mathbf{k}_\mu$ is a null vector satisfying $\mathbf{k}_\mu
\mathbf{k}^\mu=0$ and $\eta$ is energy density of the pure
radiation (null dust). Using this property  we can easily
construct the Kasner generalization of radiating Levi-Civita
solution. For the functions $K_0,U_0,W_0$ we will use the
functions $K, U, W$ of LCK solutions, namely have the metric:
\begin{eqnarray}\label{lckr}
&&K=F(r-t)+k^2\ln{(c_1r+c_2)}+q^2\ln{(c_3t+c_4)} \nonumber\\
&&U=k\ln{(c_1r+c_2)}+q\ln{(c_3t+c_4)}\nonumber \\
&&W=\alpha (c_1r+c_2)(c_3t+c_4) \\
&&q=k+\epsilon,\ \epsilon=\pm 1,\nonumber
\end{eqnarray}
which are solutions of (\ref{emt}) with the energy density:
\begin{eqnarray}\label{edense1}
&&\eta=\frac{(c_2c_3-c_1(c_4+c_3(t-r)))\,\dot{F}}{(c_1r+c_2)(c_3t+c_4)}.
\end{eqnarray}
Notice that both $c_1$ and $c_2$ cannot vanish simultaneously.
This is also true for $c_3$ and $c_4$. When $F=const.$ this
solution reduces to the LCK spacetime. Also, when we take $c_3=0$
$c_4\neq 0$ we get the radiating Levi-Civita solution of the form:
\begin{equation}\label{lcr1}
ds^2=e^{2F}r^{2(k^2-k)}(-dt^2+dr^2)+r^{2k}dz^2+\alpha^2r^{2(1-k)}d\phi^2,
\end{equation}
The energy density is
\begin{equation}
\eta=-\frac{\dot{F}}{r}
\end{equation}
To have a positive energy density, here we need $\dot{F}<0$. Also
for $c_1=0$ $c_2\neq 0$ we get the radiating Kasner solution with
the metric:
\begin{equation}
ds^2=e^{2F(t-r)}t^{2(k^2-k)}(-dt^2+dr^2)+t^{2k}dz^2+t^{2(1-k)}d\phi^2.
\end{equation}
For this radiating Kasner solution it is better to think the
coordinates as the Cartesian coordinates. This metric describes a
pure radiation moving in the $r$ direction in the Kasner
spacetime. The energy density is the negative of the Levi-Civita
case and $\dot{F}$ must be positive in order to have positive
energy density since for this case
\begin{equation}
\eta=\frac{\dot{F}}{t}.
\end{equation}
 At $t=0$ this metric has a Kasner type cosmological singularity
except $k=0$ and $k=1$.

If we have
\begin{equation}
c_2c_3-c_1c_4\ge 0,\ c_1 c_3 (r-t)\dot{F} >0,\ \dot{F} >0,
\end{equation}
or
\begin{equation}
c_2c_3-c_1c_4\le 0,\ c_1 c_3 (r-t)\dot{F} <0,\ \dot{F} <0,
\end{equation}
in (\ref{edense1}) then the energy density of the solution
(\ref{lckr}) is positive.

For example the following choice
\begin{equation}
c_2=c_4=0,\ c_1=c_3=1,\ F=-a(t-r)^n,\ n=1,2,3...,
\end{equation}
where $a>0$ is a constant leads to positive energy solutions when
$n$ is even.

This spacetime (\ref{lckr}) contains in general a Kasner type
cosmological singularity at $t=0$ and also it is singular at the
axis (We take $c_2=c_4=0$ in (\ref{lckr})). The spacetime is not
singular for the particular values of the parameters $\epsilon=1$,
$k=0$ and $\epsilon=-1$, $k=1$. The cosmological singularity seems
to unavoidable but if one is able to find a regular interior
radiating solution containing the symmetry axis, then we can avoid
having a line singularity at $r=0$ since our solution could be  an
exterior solution of a radiating nonstatic cylindrical source. The
spacetime is well behaved for $t>0$ and $r>0$.
\subsection{Some
properties of the solution}
\subsubsection{NP coefficients}
Here we analyze 
Ricci and Weyl scalars of the metric (\ref{lckr}) using a null
tetrad.  For $\epsilon=1$ we have the spin coefficients:
\begin{eqnarray}
&&\Phi_{00}=\frac{(t-r)F'}{e^{2F\,r^{2k^2-2k+1}\,t^{2k^2+2k+1}}}\\
&&\Psi_{0}=\big(k((1+k)(1+2k)r^2-4(k^2-1)rt+(k-1)(2k-1)t^2)
\\&&\phantom{aa}-((1+2k)r+(1-2k)t)2rtF'\big)/(2e^{2F}\,r^{2k^2-2k+2}\,t^{2k^2+2k+2})\\
&&\Psi_{2}=\frac{k((1+k)r^2+(1-k)t^2)}{2e^{2F}\,r^{2k^2-2k+2}\,t^{2k^2+2k+2}}\\
&&\Psi_{4}=\frac{k((1+k)(1+2k)r^2+4(k^2-1)rt+(k-1)(2k-1)t^2}{2e^{2F}\,r^{2k^2-2k+2}\,t^{2k^2+2k+2}}.
\end{eqnarray}
This shows that only for $\epsilon=1$ case, for $k=0$, $\Psi_2$
and $\Psi_4$ vanishes and the spacetime is Petrov type D. For
other values of $k$, $\Psi_0$, $\Psi_2$ and $\Psi_4$ is
nonvanishing and Petrov type is I. For $\epsilon=-1$ we have:
\begin{eqnarray}
&&\Phi_{00}=\frac{(t-r)F'}{e^{2F}\, r^{k^2-2k+1}\,t^{2k^2-6k+5}}\\
&&\Psi_{0}=\big((k-1)((k-2)(2k-3)r^2-4k(k-2)rt+k(2k-1)t^2)\\
&&\phantom{aaa}-((2k-3)r+(1-2k)t)2rtF'\big)/\big(2\,e^{2F}\,
r^{2k^2-2k+2}\, t^{2k^2-6k+6}\big)\\
&&\Psi_{2}=\frac{(k-1)(k-2)r^2-kt^2}{2\,e^{2F}\, r^{2k^2-2k+2}\,
t^{2k^2-6k+6}}\\
&&\Psi_{4}=\frac{(k-1)(k-2)(2k-3)r^2+4k(k-2)rt+k(2k-1)t^2}{2\,e^{2F}\,
r^{2k^2-2k+2}\, t^{2k^2-6k+6}}.
\end{eqnarray}
For the $\epsilon=-1$ case, $\Psi_0$, $\Psi_2$ and $\Psi_4$ is
nonvanishing and the spacetime is Petrov type I except for $k=1$
where $\Psi_2$ and $\Psi_4$ is vanishing and the spacetime is
Petrov type N.
\subsubsection{Radial acceleration of test
particles} The radial acceleration of a test particle initially at
rest in a constant radius in the spacetime (\ref{lckr}) is given
by:
\begin{equation}\label{radacc}
\ddot{r}=\frac{(k-k^2)r^{-1}-F'}{e^{F}r^{k^2-k}\,t^{q^2-q}}.
\end{equation}
If we compare (\ref{radacc}) with the LCK metric, we see that the
main difference is the term $\sim F'$ which characterizes the null
radiation. When the $F'$ is positive, the axis is more attractive
whereas when it is negative, the axis is less attractive. Thus,
the presence of null dust may  alter the particle motion.
\subsubsection{Circular geodesics}
Let us study the equations of a test particle following a circular
geodesics in the spacetime (\ref{lckr}). Let us denote the angular
velocity of a particle moving along a geodesics as $w$, then we
have:
\begin{equation}
\omega^2=\frac{(k^2-k)r^{-1}+F'}{(1-k)e^{2F}r^{2k^2-1}t^{2(q^2-1)}},
\end{equation}
which results
\begin{eqnarray}
\left(\frac{ds}{dt}\right)^2=\left(\frac{k^2-k+rF'}{1-k}-1\right)e^{2F}\,r^{2(k^2-2)}\,
t^{2(q^2-q)}.
\end{eqnarray}
Thus, the circular geodesics are timelike if the expression inside
the parentheses is negative, null if it is zero and spacelike if
it is positive. For the Levi-Civita and LCK metrics the ranges of
$k$ where the geodesics are timelike, spacelike or null are the
same. However, here we have extra terms proportional to $r\, F'$
and they are in general depends on time and the radial coordinate.
 This might
have some consequences on particle motion. For example, when time
passes, a particle following a circular geodesics may not continue
to its motion since such geodesics become spacelike. Also for a
given $k$, the circular geodesics might be restricted to a certain
radius. Hence, The presence of the null radiation clearly affects
the dependence of these ranges to the parameter $k$.
\subsection{A Radiating nonstatic string-like object}
Using the property of the Einstein-Rosen type solutions, we can
construct examples of interior solutions having a nonstatic
radiating object with a cosmic string like equation of state and
generating outer radiating spacetime for particular values of the
parameters $k$ and $q$ . The interior and exterior metrics are
given by:
\begin{eqnarray}
&&ds^2_{-}=t^{4}\left(e^{2F(r-t)}(-dt^2+dr^2)+A(r)^2\,d\phi^2\right)+t^{-2}dz^2,\\
&&ds^2_{+}=t^{4}\left(e^{2F(r-t)}(-dt^2+dr^2)+\alpha^2
r^2\,d\phi^2\right)+t^{-2}dz^2,
\end{eqnarray}
with the energy momentum tensor:
\begin{eqnarray}
&&T_{\mu\nu-}=T_{\mu\nu-}^{(R)}+T_{\mu\nu-}^{(S)},\quad T_{\mu\nu+}=\eta_{+}\mathbf{k}_\mu \mathbf{k}_\nu, \\
&&T_{\mu\nu-}^{(R)}=\kappa\,\eta_{-} \mathbf{k}_\mu
\mathbf{k}_\nu,\
k_\mu=(1,1,0,0) \\
&&T^{0\ (S)}_{\phantom{0}0-}=T^{z\ (S)
}_{\phantom{0}z-}=-\kappa\,\mu\\
 &&\eta_{-}=\frac{\left(t\,A' -A\right) F'}{
 t\,A},\quad \mu=\frac{-A''}{A\,e^{2F} t^{-4}},\quad \eta_+= \frac{\left(t -r\right) F'}{
 t\,r}
\end{eqnarray}

In these solutions, the interior and exterior metrics can be
smoothly matched if the metrics and their first derivatives are
continuous on the boundary of the stringlike object. Since we have
chosen same inner and outer coordinates, this can be fulfilled if
$A(r_0)=\alpha r_0 $ and $A'(r_0)=\alpha$. These are called
Lichnerowicz boundary conditions \cite{Lichnerowicz} and can be
satisfied for the present case easily. For example if we choose
$A(r)=\sin(b\,r )$ then the junction conditions yield:
\begin{equation}
\alpha=\sin(b r_0),\quad r_0=\tan(b\, r_0)/b
\end{equation}
which can be easily satisfied since we have more parameters than
equations.

 Here the problem of these solutions is that, unlike
$\mu$, it seems that it may be impossible to $\eta_{-}$ to be
positive for all ranges of $r$ and $t$. However, we can avoid
negative energy density if we limit the $r$ and $t$ ranges with
limited values where $\eta_- >0$. %
Then, the solution can represent a radiating nonstationary cosmic
string like object emitting null radiation.
\section{A Stiff Fluid of generalized Kasner form }

Let us consider the following metric:
\begin{equation}
ds^2=r^{2(k^2-k)}\, t^{2(q^2-q+a)}(-dt^2+dr^2) 
+ r^{2k}\, t^{2q}dz^2 +\ P^2\, r^{2(1-k)}\, t^{2(1-q)}d\phi^2,
\end{equation}
which deviates from LCK solution by a parameter $a$. For
$a=-k^2-q^2+2kq+1$ we have a nonstatic  stiff fluid with the
equation of state
\begin{equation}
-G^{0}_{\phantom{0}0}=G^{r}_{\phantom{0}r}=G^{z}_{\phantom{0}z}=G^{\phi}_{\phantom{0}\phi}=a
\ r^{2(k-k^2)}t^{2(q-q^2-a-1)}.
\end{equation}
When $a \rightarrow0$ we recover the LCK  solution. This solution
has the similar singularity behavior as the LCK metric and it is
singular in general at $r=0$ and $t=0$. For a special values of
$k,q$ we can avoid the singularity at the axis. For example for
$k=q=0$ we have $a=1$ and the metric becomes
\begin{equation}
ds^2=t^2(-dt^2+dr^2+r^2d\phi^2)+dz^2,
\end{equation}
with
\begin{equation}
-G^{0}_{\phantom{0}0}=G^{r}_{\phantom{0}r}=G^{z}_{\phantom{0}z}=G^{\phi}_{\phantom{0}\phi}=
t^{-4}.
\end{equation}
This metric describes a cosmological solution where at $t=0$ we
have a big bang singularity, the Kretchman scalar is $K\sim t ^8$,
then we have an universe filled with an isotropic stiff fluid with
the equation of state $\rho=p$. Since the energy density goes with
$t^{-4}$, for large $t$ the it becomes negligible  at late times
and practically at ($t\rightarrow\infty$) we get vacuum universe.
\section{Conclusions}
In this paper we have first investigated some physical properties
of the nonstatic vacuum solutions in cylindrical coordinates with
Kasner type  time dependence. They can describe the exterior
regions of nonstatic line sources and nonstatic straight strings
\cite{Schabes}-\cite{Sen} having nonvanishing gravitational
potential. For each constant time slice they reduce to the
Levi-Civita metric. For each Levi-Civita parameter, $s$, there are
in general two corresponding nonstatic vacuum solution of this
form depending on $\epsilon=\pm 1.$ We have studied some physical
properties of this space-time and compared with the static
Levi-Civita spacetime. This metric is in the form of generalized
Kasner solutions studied before \cite{Mcintosh}. Also, by a
coordinate transformation, it reduces to Einstein-Rosen soliton
waves \cite{Carmeli}.  We have discovered some differences and
similarities between LC and LCK spacetimes. We believe that the
form of the metric (\ref{lckGN}) is suitable for future
applications.

Next, we generalized the discussion to a cylindrical nonstatic
metric corresponding to an exterior atmosphere of a cylindrical
radiating nonstatic source having generalized Kasner type metric.
The atmosphere has an outgoing radial pure radiation as well as
incoming and outgoing gravitational radiation. For some special
cases of our parameter $k$, these solutions reduce to the exterior
field of the radiating nonstatic cosmic string-like objects.

Finally, we have presented a stiff fluid solution by a small
deviation of $g_{tt}$ component of the metric from LCK spacetime.
This solution is also nonstatic and nonstationary and it is
another sign off richness of cylindrically symmetric sources of
general relativity together with the previous solutions we have
discussed.

\section*{Acknowledgments}

I would like to thank Metin Ar\i k for reading the manuscript and
useful discussions. Some portions of this work is done with the
help of {\it GRTensor II} \cite{grtensor}.

\end{document}